\documentclass[lettersize,journal]{IEEEtran}
\usepackage{amsmath,amsfonts}
\usepackage{array}
\usepackage[caption=false,font=normalsize,labelfont=sf,textfont=sf]{subfig}
\usepackage{textcomp}
\usepackage{stfloats}
\usepackage{url}
\usepackage{verbatim}
\usepackage{graphicx}
\usepackage{cite}
\usepackage{color}
\usepackage{comment}
\usepackage{multirow}
\usepackage[english]{babel}
\usepackage[utf8x]{inputenc}
\usepackage[T1]{fontenc}
\usepackage{tikz}
\usetikzlibrary{positioning, arrows.meta, fit, backgrounds, calc, decorations.pathreplacing}
\usepackage{xcolor}
\graphicspath{{images/}}
\usepackage{adjustbox}
\usepackage{booktabs}
\usepackage{float}
\usepackage[most]{tcolorbox}
\usepackage{balance}

\def\BibTeX{{\rm B\kern-.05em{\sc i\kern-.025em b}\kern-.08em
    T\kern-.1667em\lower.7ex\hbox{E}\kern-.125emX}}

\newcommand{\ea}{\textit{et al.}}

\newcommand{\smallsection}[1]{\noindent\textbf{#1.}}

\begin{document}

\title{Shard the Gradient, Scale the Model: Serverless Federated Aggregation via Gradient Partitioning}

\author{Amine~Barrak
\thanks{A. Barrak is with the Department of Computer Science and Engineering, Oakland University, Rochester, MI 48309, USA (e-mail: aminebarrak@oakland.edu).}%
\thanks{Manuscript received \today.}%
}

\markboth{IEEE Transactions on Cloud Computing}%
{Barrak: Serverless Federated Aggregation via Gradient Partitioning}

\maketitle

Federated learning (FL) aggregation on serverless platforms faces a hard scalability ceiling: existing architectures ($\lambda$-FL, LIFL) partition \emph{clients} across aggregators, but every aggregator must hold the complete model gradient in memory. When gradients exceed the per-function memory limit (e.g., 10\,GB on AWS Lambda), aggregation becomes infeasible regardless of tree depth or branching factor. We propose \textsc{GradsSharding}, which instead partitions the \emph{gradient tensor} into $M$ shards, each averaged independently by a serverless function that receives contributions from all clients. Because FedAvg averaging is element-wise, this produces bit-identical results to tree-based approaches, so model accuracy is invariant by construction. Per-function memory is bounded at $\mathcal{O}(|\theta|/M)$, independent of client count, enabling aggregation of arbitrarily large models. We evaluate \textsc{GradsSharding} against $\lambda$-FL and LIFL through HPC experiments and real AWS Lambda deployments across model sizes from 43\,MB to 5\,GB. Results show a cost crossover at $\sim$500\,MB gradient size, 2.7$\times$ cost reduction at VGG-16 scale, and that \textsc{GradsSharding} is the only architecture that remains deployable beyond the serverless memory ceiling.


\section{Introduction}
\label{sec:introduction}

Federated learning (FL)~\cite{mcmahan2017communication} trains machine learning models across distributed data silos without centralizing raw data. In the canonical FedAvg protocol, a central parameter server distributes a global model, each participating client trains locally and returns a gradient or weight delta, and the server averages the updates to produce the next global model~\cite{li2014scaling}. As model sizes grow from hundreds of millions to tens of billions of parameters and client populations scale into the hundreds per round, the aggregation step, which must buffer and combine these updates in memory, becomes the primary systems bottleneck.

The models being trained in FL deployments have grown dramatically in recent years. Early FL benchmarks focused on lightweight architectures such as ResNet-18 (43\,MB gradient footprint) or simple CNNs. Modern applications, however, increasingly involve large vision models like VGG-16 (512\,MB) and transformer-based architectures such as GPT-2 Large (2\,GB) or larger language models exceeding 10\,GB in parameter size~\cite{brown2020language,touvron2023llama}. This rapid growth in model scale creates a fundamental challenge for the aggregation infrastructure: the server must receive, buffer, and combine gradient tensors that can individually exceed several gigabytes, from potentially hundreds of clients per round.

A key observation about FL workloads is that the aggregation phase is short-lived relative to client-side training: each round concentrates server-side work into brief, bursty windows separated by long idle periods while clients train locally. Despite this being widely cited as motivation for serverless FL~\cite{jayaram2022lambda,ciobotaru2024lifl, jayaram2022adafed}, no prior work has empirically quantified the idle ratio across model architectures of varying scale. In a standard FedAvg deployment with $N$ clients training for $E$ local epochs, the parameter server (PS) must wait for all clients to complete training before it can aggregate. If this idle fraction exceeds 90\% of each round, provisioning a persistent server is wasteful: the infrastructure sits unused for the vast majority of the training process. Serverless platforms (AWS Lambda, Azure Functions, Google Cloud Functions) offer a natural alternative, providing automatic scaling, fine-grained billing, and zero idle cost. However, serverless functions operate under strict resource constraints: AWS Lambda, for instance, limits individual function memory to 10\,GB and execution time to 15 minutes. For small models like ResNet-18, these limits are easily accommodated. But as model sizes approach and exceed the gigabyte range, a single serverless function can no longer hold the full gradient tensor required for aggregation, which makes existing aggregation architectures infeasible on serverless infrastructure.

Existing serverless FL systems have explored two primary strategies to work within these limits. $\lambda$-FL~\cite{jayaram2022lambda} adopts a \emph{tree-based} aggregation topology inspired by MapReduce: leaf aggregators each handle a subset of $k$ clients, and intermediate aggregators combine partial results upward. This reduces the number of clients each aggregator serves, but every aggregator still must hold the entire gradient tensor in memory, so per-function memory remains proportional to the full model size $|\theta|$. LIFL~\cite{ciobotaru2024lifl} takes a different approach by colocating aggregation functions on physical nodes with shared memory, eliminating data serialization overhead and enabling zero-copy gradient access. While highly efficient when functions share a node, LIFL's memory advantages depend on locality-aware placement and, like $\lambda$-FL, each aggregator must hold the complete model gradient in memory. For both systems, when the gradient tensor exceeds the serverless memory limit (10\,GB on AWS Lambda), aggregation becomes fundamentally impossible, regardless of how many clients or tree levels are used.

The root cause is that both approaches split the work by clients, each aggregator receives updates from fewer clients, but still processes the entire gradient tensor. We propose \textsc{GradsSharding}, which partitions the model dimension instead: rather than each aggregator handling the full gradient from fewer clients, each aggregator handles a slice of the gradient from all clients. We adapt this to the serverless context: S3 replaces the persistent parameter store, and stateless Lambda functions replace the server nodes. In \textsc{GradsSharding}, each client's gradient tensor is split into $M$ shards along the parameter axis. For each shard index $j$, an independent serverless function collects the $j$-th shard from \emph{all} $N$ clients, computes the element-wise average, and returns the result. Because FedAvg computes a simple element-wise average, averaging each shard separately and concatenating the results produces exactly the same output as averaging the full gradient at once. This means model accuracy and convergence are completely unaffected. The only difference between the approaches is how the aggregation work is organized.

The key advantage is \textbf{memory-bounded aggregation}: each function only needs $\mathcal{O}(|\theta|/M)$ memory, regardless of the full model size or the number of clients $N$. If a model is too large for a single function, we simply increase $M$ to make each shard smaller, for example, staying within AWS Lambda's 10\,GB limit. This means \textsc{GradsSharding} can handle arbitrarily large models, while $\lambda$-FL and LIFL break down once the gradient exceeds what a single function can hold.

This paper makes the following contributions:

\begin{enumerate}
    \item We propose \textsc{GradsSharding}, the first serverless FL aggregation architecture to decouple per-function memory from both model size and client count by sharding the gradient tensor across $M$ parallel functions, each bounded at $\mathcal{O}(|\theta|/M)$.

    \item We compare \textsc{GradsSharding} against the state-of-the-art serverless FL architectures $\lambda$-FL and LIFL under identical training configurations, providing an end-to-end evaluation across memory footprint, aggregation latency, and cost.

    \item We conduct a benchmark study of \textsc{GradsSharding} on real AWS Lambda infrastructure with a full shard sweep ($M \in \{1,2,4,8,16\}$) on VGG-16, providing ground-truth measurements of S3 throughput, parallel execution speedup, and actual AWS billing.

    \item We deploy all three architectures on real AWS Lambda across four model scales (43\,MB to 5\,GB), measuring end-to-end aggregation latency and cost. The results identify the crossover point where \textsc{GradsSharding} becomes cheaper than tree-based approaches and the model size beyond which full-gradient architectures exceed Lambda's memory limit.
\end{enumerate}

\noindent A replication package containing all source code, experiment scripts, raw results, and figure generation scripts is available at: \footnote{https://github.com/AmineBarrak/Serverless-aggregation-grads-sharding}.
\section{Related Work}
\label{sec:related}

This section reviews the literature along two dimensions directly relevant to our contribution: federated learning aggregation strategies that address scalability bottlenecks, and serverless computing systems designed for distributed machine learning workloads.

\subsection{Federated Learning Aggregation}

McMahan~\ea~\cite{mcmahan2017communication} introduced FedAvg, the foundational synchronous aggregation protocol where a central server averages client model updates each round. Their contribution established that averaging locally trained models can match centralized training accuracy while dramatically reducing communication rounds.

Hierarchical aggregation addresses the single-server bottleneck by inserting intermediate aggregators between clients and the global server. Liu~\ea's HierFAVG~\cite{liu2020client} introduces edge servers as this intermediate tier, reducing wide-area communication by up to 5$\times$ while maintaining convergence under non-IID data. FedAT~\cite{chai2021fedat} combines asynchronous inter-group aggregation with synchronous intra-group updates, achieving up to 3.5$\times$ faster convergence by masking stragglers across tiers. Both designs still assume persistent aggregator nodes that hold the full gradient, so they do not address the per-function memory cap we target.

A parallel line of work reduces the \emph{volume} of each update rather than restructuring the aggregation topology. QSGD~\cite{alistarh2017qsgd} quantizes gradients to few-bit encodings with provable convergence, cutting per-round communication by up to 8$\times$. Top-$k$ sparsification~\cite{aji2017sparse} transmits only the largest 1\% of gradient entries at comparable accuracy. Count-Sketch compression~\cite{ivkin2019communication} achieves sublinear $\mathcal{O}(1/\epsilon^2)$ communication complexity independent of model dimension. These techniques are orthogonal to \textsc{GradsSharding} and can be composed by compressing each shard before upload.

Decentralized and peer-to-peer designs sidestep the central aggregator entirely. Heged{\H{u}}s~\ea~\cite{hegedus2021decentralized} show that gossip protocols match FedAvg on IID data but degrade under non-IID heterogeneity because of inconsistent averaging across the peer graph. Lalitha~\ea~\cite{lalitha2019peer} propose a Bayesian peer-to-peer FL framework that handles heterogeneity well but requires persistent peer connections, which are incompatible with ephemeral serverless functions. FedProx~\cite{li2020federated} generalizes FedAvg with a proximal term for systems and statistical heterogeneity, but does not address the memory constraint of the aggregation step itself.

In the domain of model parallelism, Shoeybi~\ea~\cite{shoeybi2019megatron} proposed Megatron-LM, which partitions transformer layers across GPUs using intra-layer tensor parallelism. Their contribution demonstrated that splitting individual matrix multiplications across devices enables training models with billions of parameters beyond single-GPU memory limits. Zhao~\ea~\cite{zhao2023pytorch} contributed PyTorch FSDP (Fully Sharded Data Parallel), which shards model parameters, gradients, and optimizer states across data-parallel workers, materializing full parameters only during forward and backward computation.

\textsc{GradsSharding} applies a similar sharding principle but in a fundamentally different context: rather than sharding for distributed \emph{training} across persistent GPU workers, we shard for distributed \emph{aggregation} on stateless, memory-constrained serverless functions, requiring no inter-shard communication.

\subsection{Serverless Computing for Machine Learning}

Serverless computing has been extensively adopted across ML workflows, spanning training, inference, and data preprocessing, as surveyed by Barrak~\ea~\cite{barrak2022serverless}. Jonas~\ea~\cite{jonas2017occupy} introduced PyWren, one of the earliest systems to demonstrate that serverless functions (AWS Lambda) can execute general-purpose parallel computations at scale. Their contribution established the pattern of using cloud object storage (S3) as the communication substrate between stateless function instances, a design pattern that \textsc{GradsSharding} adopts for shard transfer. Carreira~\ea~\cite{carreira2019cirrus} proposed Cirrus, a serverless framework for end-to-end ML workflows including distributed training on AWS Lambda. Their contribution identified cold start latency and limited function memory as the primary bottlenecks for serverless ML, reporting up to 10$\times$ overhead compared to dedicated cluster deployments for iterative training workloads.

Several studies have explored distributed training over serverless computing \cite{barrak2023exploring}. Jiang~\ea~\cite{jiang2021towards} provided a detailed performance analysis showing that communication overhead between stateless functions dominates total execution time, accounting for 60--80\% of wall-clock time depending on model size, motivating the need for communication-efficient aggregation designs. Barrak~\ea~\cite{barrak2023spirt} proposed SPIRT, a peer-to-peer serverless training architecture that eliminates the central aggregator entirely by having serverless functions coordinate directly via object storage, with built-in fault tolerance through gradient checkpointing and peer replacement. Unlike the centralized aggregation systems studied in this paper, SPIRT targets the training phase itself rather than the aggregation topology. Shankar~\ea~\cite{shankar2020serverless} proposed a cost model for serverless ML inference, contributing analytical expressions for the tradeoff between provisioned capacity and pay-per-invocation billing. While focused on inference rather than training, their cost modeling methodology informed our AWS cost analysis for \textsc{GradsSharding}.

Jayaram~\ea's $\lambda$-FL~\cite{jayaram2022lambda} adapts tree-based aggregation to serverless execution, organizing client updates into leaf aggregators that compute partial averages before a root aggregator combines the results. The design exploits serverless pay-per-use pricing for intermittent FL workloads. However, every aggregator in the tree must hold the full gradient tensor in memory, so the architecture becomes infeasible when the gradient exceeds the serverless memory limit.

Ciobotaru~\ea~\cite{ciobotaru2024lifl} proposed LIFL, a lightweight event-driven serverless platform designed for FL workloads. LIFL exploits shared memory for zero-copy data transfer between colocated functions and locality-aware placement to maximize colocation, achieving significant speedups over $\lambda$-FL. However, its performance advantages depend on functions sharing a physical node. On serverless platforms like AWS Lambda, where each function runs in an isolated container, shared memory is unavailable and all communication falls back to network-based transfer.

Grafberger~\ea~\cite{grafberger2021fedless} proposed FedLess, a serverless FL framework built on Apache OpenWhisk that automates client function deployment and supports multiple aggregation strategies. Their contribution demonstrated that serverless FL can reduce operational complexity by eliminating client infrastructure management, but did not address the per-function memory constraints that limit model size during aggregation. Kim~\ea~\cite{kim2022serverless} contributed an empirical study of serverless FL on heterogeneous edge devices, showing that cold start variability across device types introduces up to 4$\times$ latency variance between rounds, motivating the need for aggregation designs that are robust to function startup delays.

Table~\ref{tab:comparison} positions \textsc{GradsSharding} relative to the two most closely related serverless FL systems. Our approach is, to our knowledge, the first to bound per-function memory independently of client count by partitioning the gradient tensor rather than the client population.

\begin{table}[t]
\centering
\caption{Comparison of serverless FL aggregation approaches. $N$: clients per round, $|\theta|$: model parameters, $M$: number of gradient shards, $k$: clients per leaf aggregator, $L_1$: level-1 aggregator count.}
\label{tab:comparison}
\resizebox{\columnwidth}{!}{%
\small
\begin{tabular}{@{}lccc@{}}
\toprule
\textbf{Property} & \textbf{$\lambda$-FL} & \textbf{LIFL} & \textbf{GradsSharding} \\
\midrule
Partition axis & Clients & Clients & Parameters \\
Mem.\ per fn & $\mathcal{O}(|\theta|)$\textsuperscript{\dag} & $\mathcal{O}(|\theta|)$\textsuperscript{*} & $\mathcal{O}(|\theta|/M)$ \\
Depends on $N$ & Yes & Yes & No \\
Load per agg & $k \!\times\! |\theta|$ & $\lceil N/L_1\rceil \!\times\! |\theta|$ & $N \!\times\! |\theta|/M$ \\
Inter-agg comms & Kafka MQ & Shared mem & None \\
Infra required & Msg queue & Coloc.\ nodes & Object store \\
Parallelism & $\lceil N/k\rceil$ & Per-node & $M$ shards \\
\bottomrule
\multicolumn{4}{@{}l@{}}{\textsuperscript{\dag}\footnotesize With streaming; $\mathcal{O}(\sqrt{N}|\theta|)$ if buffered.}\\
\multicolumn{4}{@{}l@{}}{\textsuperscript{*}\footnotesize Amortized via shared memory; physical memory shared across node.}
\end{tabular}%
}
\end{table}

\section{Study Design}
\label{sec:design}

This section presents the three system architectures under evaluation, followed by the research questions that guide our empirical study.

\subsection{System Architectures}

\begin{figure*}[t]
\centering
\input{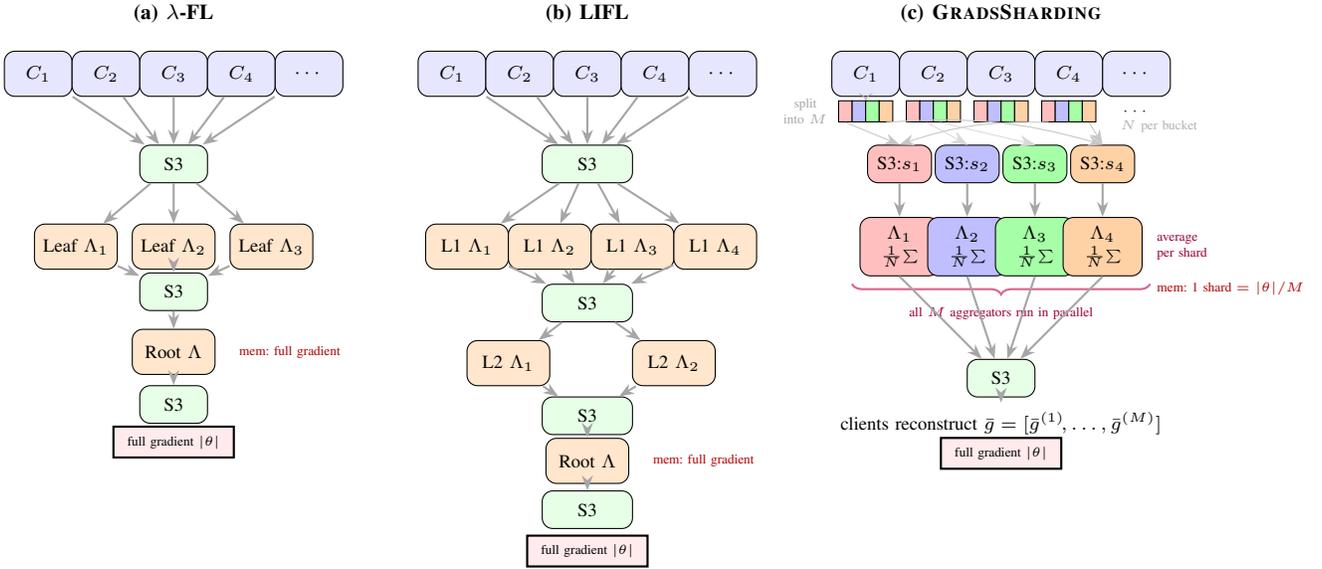}
\caption{The three serverless FL aggregation architectures (schematic; actual aggregator counts depend on $N$ and branching factor). Orange boxes ($\Lambda$) are Lambda functions; green boxes are S3 buckets. In all three architectures, the final aggregated gradient is written back to S3. (a)~$\lambda$-FL: two-level tree, each Lambda holds the full gradient. (b)~LIFL: three-level hierarchy, all inter-level communication via S3, each Lambda holds the full gradient. (c)~\textsc{GradsSharding}: $M$ parallel Lambdas each average one shard across all $N$ clients; each holds only $|\theta|/M$ in memory.}
\label{fig:architectures}
\end{figure*}

All three systems implement the FedAvg~\cite{mcmahan2017communication} aggregation protocol: in each round, the server distributes the current global model, selected clients perform local training and return gradient updates, and the server aggregates these updates via weighted averaging. The systems differ in \emph{how} the aggregation step is organized and executed on serverless infrastructure. Figure~\ref{fig:architectures} illustrates the three architectures side by side.

\subsubsection{$\lambda$-FL: Tree-Based Aggregation}

$\lambda$-FL~\cite{jayaram2022lambda} organizes aggregation as a two-level tree. Given $N$ client updates per round, the system computes $k = \max(2, \lceil\sqrt{N}\rceil)$ and partitions clients into $\lceil N/k \rceil$ groups. Each group is assigned to a \emph{leaf aggregator} (a serverless function that computes the partial average of its assigned gradients). A \emph{root aggregator} then combines all leaf outputs into the final aggregated gradient.

Client gradients are uploaded to S3, and leaf aggregators are triggered via S3 event notifications when all gradients for their group are available. Each leaf reads its assigned gradients, computes a streaming average, and writes its partial result back to S3. The root aggregator is then triggered, reads all $\lceil N/k \rceil$ leaf outputs from S3, combines them into the final aggregated gradient, and writes the result back to S3.

The tree topology distributes aggregation work across leaves, with each leaf handling at most $k$ clients' gradients. However, every aggregator (both leaf and root) must hold the \emph{full model gradient} in memory for averaging. Each aggregator requires two full-gradient-sized buffers: one for the incoming gradient and one for the running average. This means that when the full gradient no longer fits in a single function's memory, the tree structure cannot help: the architecture becomes infeasible regardless of tree depth or branching factor.

\subsubsection{LIFL: Shared-Memory Event-Driven Aggregation}

LIFL~\cite{ciobotaru2024lifl} redesigns the serverless aggregation pipeline around three principles: (1) lightweight event-driven function execution with minimal cold-start overhead, (2) shared memory for zero-copy data transfer between colocated functions, and (3) locality-aware placement to maximize colocation.

The system deploys a \emph{gateway} on each worker node that receives incoming model updates and writes them to a node-local shared memory store. Aggregator functions on the same node read directly from shared memory, avoiding network transfer and serialization. A \emph{hierarchy-aware autoscaler} manages the aggregator pool, promoting frequently-used aggregators and reusing warm instances to minimize cold starts.

LIFL uses a multi-level hierarchical tree (typically 3 levels) where level-1 aggregators each process a subset of clients, level-2 aggregators combine level-1 results, and a root aggregator produces the final output and writes it back to S3. In its original design, colocated aggregators communicate via shared memory; however, on serverless platforms such as AWS Lambda, each function runs in an isolated container with no shared address space (Section~\ref{sec:setup}). On Lambda, all inter-aggregator communication therefore uses S3: each level writes its partial results to S3, and the next level reads them. Like $\lambda$-FL, each aggregator at every level must hold the full gradient tensor in memory, making LIFL subject to the same memory ceiling: when the model gradient exceeds the serverless function memory limit, aggregation is impossible.

\subsubsection{\textsc{GradsSharding}: Gradient Tensor Partitioning}

\textsc{GradsSharding} takes a fundamentally different approach by partitioning the \emph{gradient tensor} rather than the \emph{client population}. The aggregation pipeline for a single round proceeds in four steps:

\smallsection{Step 1 -- Shard} Each client's gradient vector $g_i \in \mathbb{R}^{|\theta|}$ is split into $M$ contiguous shards: $g_i = [g_i^{(1)}, g_i^{(2)}, \ldots, g_i^{(M)}]$ where $g_i^{(j)} \in \mathbb{R}^{|\theta|/M}$.

\smallsection{Step 2 -- Upload} All shards are uploaded to S3, organized by round and shard index. Each client performs $M$ parallel PUT operations.

\smallsection{Step 3 -- Aggregate} $M$ independent serverless functions are triggered via S3 event notifications when all $N$ client shards for their index are available. Function $j$ downloads shard $j$ from all $N$ clients using streaming accumulation, computes the element-wise average $\bar{g}^{(j)} = \frac{1}{N}\sum_{i=1}^{N} g_i^{(j)}$, and writes the averaged shard back to S3.

\smallsection{Step 4 -- Reconstruct} Each client reads all $M$ averaged shards from S3 and concatenates them to reconstruct the full aggregated gradient $\bar{g} = [\bar{g}^{(1)}, \ldots, \bar{g}^{(M)}]$. The client then updates its local model for the next round.

\smallsection{Memory analysis} Each shard aggregator $j$ uses streaming accumulation: it maintains a running average in memory and processes one client's shard at a time, requiring only two buffers, each of size $|\theta|/M$, one for the running average and one for the incoming shard. Memory therefore scales as $\mathcal{O}(|\theta|/M)$, independent of the number of clients $N$. By choosing $M$ large enough that a single shard fits within the serverless memory budget, any model can be aggregated regardless of its total size. The same streaming approach applies to $\lambda$-FL and LIFL, but their aggregators each require two full-gradient-sized buffers ($\mathcal{O}(|\theta|)$); our RQ3 Lambda deployment uses these streaming bounds, while the measured peaks in RQ2 reflect a simpler collect-then-average implementation.

\smallsection{Aggregation equivalence} Because FedAvg's weighted average $\bar{g} = \frac{1}{N}\sum_{i} g_i$ is element-wise, averaging contiguous coordinate blocks independently and then concatenating is algebraically identical to averaging the full vectors and then partitioning: $\frac{1}{N}\sum_{i} [g_i^{(1)}, \ldots, g_i^{(M)}] = [\frac{1}{N}\sum_{i} g_i^{(1)}, \ldots, \frac{1}{N}\sum_{i} g_i^{(M)}]$. \textsc{GradsSharding} therefore produces \emph{bit-identical} aggregated gradients to $\lambda$-FL and LIFL on every round. Accuracy and convergence are invariant by construction, so this study compares aggregation \emph{topologies}, not learning algorithms.

\smallsection{Cost model} The operational cost per round comprises Lambda compute (\$0.0000166667/GB-s) and S3 I/O (\$0.005/1{,}000 PUTs, \$0.0004/1{,}000 GETs). Table~\ref{tab:cost_model} summarizes the per-round S3 operations and memory allocation for each architecture. On AWS Lambda, all three architectures use S3 as the communication substrate, since shared memory is unavailable in Lambda's isolated containers (Section~\ref{sec:setup}). Clients write (PUT) their gradients to S3, aggregators read (GET) the gradients they need, and write their partial and final results back to S3. Clients then read the final aggregated gradient from S3 to update their local models. Note that client read-back differs across architectures: in $\lambda$-FL and LIFL, each of the $N$ clients reads a single full gradient (totaling $N$ GETs), while in \textsc{GradsSharding}, each of the $N$ clients reads $M$ averaged shards (totaling $NM$ GETs).

\begin{table}[t]
\centering
\caption{Per-round S3 operations and memory allocation for each architecture. $N$: clients per round, $|\theta|$: model parameters, $M$: gradient shards, $k$: clients per leaf, $L_1, L_2$: level-1 and level-2 aggregator counts.}
\label{tab:cost_model}
\resizebox{\columnwidth}{!}{%
\small
\begin{tabular}{@{}lccc@{}}
\toprule
 & \textbf{$\lambda$-FL} & \textbf{LIFL} & \textbf{GradsSharding} \\
\midrule
PUTs & $N + \lceil N/k \rceil + 1$ & $N + L_1 + L_2 + 1$ & $NM + M$ \\
GETs (agg) & $N + \lceil N/k \rceil$ & $N + L_1 + L_2$ & $NM$ \\
GETs (clients) & $N$ & $N$ & $NM$ \\
Mem/agg & $\mathcal{O}(|\theta|)$ & $\mathcal{O}(|\theta|)$ & $\mathcal{O}(|\theta|/M)$ \\
\# aggs & $\lceil N/k \rceil + 1$ & $L_1 + L_2 + 1$ & $M$ \\
\bottomrule
\end{tabular}%
}
\end{table}

\subsection{Research Questions}

We evaluate the serverless FL paradigm and \textsc{GradsSharding} through three research questions:

\begin{itemize}
    \item \textbf{RQ1 (Serverless Motivation):} What fraction of each FL round does the parameter server spend idle, waiting for clients to complete local training, and how does this fraction vary across model architectures?
    \item \textbf{RQ2 (Shard Ablation):} How does the number of gradient shards $M$ affect aggregation memory, latency, and cost across different model sizes?
    \item \textbf{RQ3 (Cross-Architecture Cost \& Scalability):} How do the three architectures compare in cost, latency, and feasibility on real AWS Lambda across model sizes ranging from 43\,MB to 5\,GB, and at what gradient size do full-gradient architectures exceed Lambda's memory limit?
\end{itemize}

Each research question is fully detailed---including experimental setup, approach, and results---in Section~\ref{sec:results}.
\section{Experimental Setup}
\label{sec:setup}

This section describes the shared training configuration, HPC hardware, and baseline reimplementations. Dataset, model, deployment, and cost details are described alongside each research question in Section~\ref{sec:results}.

\subsection{Training Configuration}

All experiments use the FedAvg~\cite{mcmahan2017communication} aggregation protocol with consistent hyperparameters across the three systems: learning rate $\eta = 0.01$, momentum $0.9$, and one local epoch per round. Batch size is 32 for ResNet-18 experiments and 8 for VGG-16 experiments (reduced to accommodate GPU memory with large client counts). Client selection is uniform random with full participation in each round. \textsc{GradsSharding} uses $M = 4$ shards in RQ3; RQ2 varies $M \in \{1, 2, 4, 8, 16\}$.

\subsection{Hardware Environment}

HPC experiments (RQ1 Part~A and RQ2 Part~A) are executed on the Matilda High-Performance Computing cluster at Oakland University. Each experiment runs on a compute node equipped with an NVIDIA Tesla V100 GPU (16\,GB HBM2), 8 CPU cores, and 64\,GB of system RAM.

\subsection{AWS Lambda Deployment}

Lambda experiments (RQ1 Part~B, RQ2 Part~B, and RQ3) run on AWS Lambda in us-east-1 using the Python 3.12 runtime with the AWSSDKPandas layer.

\subsection{Baseline Reimplementation}

To ensure a fair comparative analysis, we reimplemented $\lambda$-FL~\cite{jayaram2022lambda} and LIFL~\cite{ciobotaru2024lifl} within the same codebase and execution environment as \textsc{GradsSharding}. For $\lambda$-FL, we implemented the two-level tree aggregation topology with $k = \lceil\sqrt{N}\rceil$ clients per leaf and streaming accumulation within each aggregator. For LIFL, we implemented the three-level hierarchical aggregation tree with shared-memory communication between colocated aggregators, following the original design. All three systems share a common codebase for client training, data partitioning, model initialization, and metric collection. The only differences lie in the aggregation topology itself, ensuring a fair comparison.

\section{Results}
\label{sec:results}

This section presents the experimental results for each research question, organized into a consistent structure: \emph{motivation} (why we ask), \emph{approach} (how we measure), and \emph{results} (what we find).

\subsection{RQ1: Parameter Server Idle Time}

\subsubsection{Part A: HPC Measurement}

\mbox{}\\\smallsection{Motivation}
In a standard FedAvg round, the server distributes the global model, then waits for all $N$ clients to complete $E$ local epochs of training before aggregating their gradient updates. During the entire client training phase, the parameter server performs no computation, it is idle. If this idle fraction is consistently high across model scales, provisioning a persistent aggregation server is wasteful, which directly motivates the serverless paradigm.

\smallsection{Approach}
We measure the time breakdown of a single FL round for four models of increasing scale:

\begin{itemize}
    \item \textbf{Vision:} ResNet-18 (11.2M params, 43\,MB gradient) and VGG-16 (134M params, 512\,MB), trained on CIFAR-10 partitioned IID ($|D_k|=2{,}500$, $B=32$).
    \item \textbf{Language:} GPT-2 Medium (355M params, 1.4\,GB) and GPT-2 Large (774M params, 2.9\,GB), trained on synthetic token sequences ($|D_k|=2{,}500$, seq.\ len.\ 128, $B=8$).
\end{itemize}

Following McMahan~\ea~\cite{mcmahan2017communication}, we set $N=20$ clients, $E=5$ local epochs, yielding 395 training steps per client per round. For each model, we measure:

\begin{itemize}
    \item $T_{\text{train}}$: wall-clock time for one client to complete $E$ local epochs on a Tesla V100 GPU.
    \item $T_{\text{agg}}$: wall-clock time for the server to perform FedAvg aggregation (element-wise sum of $N$ gradient vectors plus scalar division) on CPU.
\end{itemize}

The PS idle ratio is computed as $T_{\text{train}} / (T_{\text{train}} + T_{\text{agg}})$, representing the fraction of each round during which the server is idle when all $N$ clients train in parallel. Since RQ1 measures the \emph{time} ratio, not model accuracy, the choice of dataset does not affect the conclusion.

\begin{figure}[h]
\centering
\includegraphics[width=0.9\columnwidth]{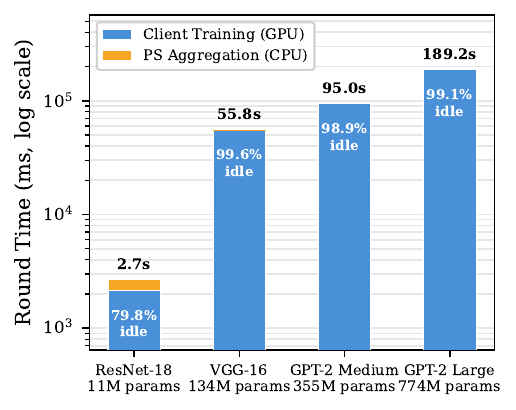}
\caption{Round time breakdown across model scales. Blue bars: client training ($T_{\text{train}}$); orange bars: server aggregation ($T_{\text{agg}}$). The PS is idle during the entire training phase.}
\label{fig:ps_idle}
\end{figure}

Table~\ref{tab:rq1_idle} reports the per-round timing breakdown and PS idle ratio for four model architectures. Figure~\ref{fig:ps_idle} visualizes the round composition.

\smallsection{Results}

\begin{table}[h]
\centering
\caption{PS idle ratio per FL round. $N=20$, $E=5$ local epochs, 395 steps/client. Training on V100 GPU; aggregation on CPU.}
\label{tab:rq1_idle}
\resizebox{\columnwidth}{!}{%
\begin{tabular}{lrrrrr}
\toprule
\textbf{Model} & \textbf{Params} & \textbf{Grad (MB)} & $T_{\text{train}}$ \textbf{(ms)} & $T_{\text{agg}}$ \textbf{(ms)} & \textbf{PS Idle (\%)} \\
\midrule
ResNet-18    & 11.2M  & 42.7  & 2{,}154  & 544 & 79.8 \\
VGG-16       & 134M   & 512   & 55{,}562  & 218 & 99.6 \\
GPT-2 Medium & 355M   & 1{,}354 & 93{,}919  & 1{,}072 & 98.9 \\
GPT-2 Large  & 774M   & 2{,}953 & 187{,}515  & 1{,}701 & 99.1 \\
\bottomrule
\end{tabular}%
}
\end{table}

\textbf{Across all four architectures, client training accounts for 79.8\%--99.6\% of each round.}
For ResNet-18, training takes 2{,}154\,ms while aggregation completes in 544\,ms, yielding a PS idle ratio of 79.8\%.
As model size increases, training time increases with model size due to larger forward and backward passes (from 2.2\,s for ResNet-18 to 187.5\,s for GPT-2 Large), while aggregation time grows more slowly (from 544\,ms to 1{,}701\,ms), reflecting the linear cost of element-wise averaging.
For VGG-16 (134M parameters), training requires 55{,}562\,ms versus only 218\,ms for aggregation, pushing the idle ratio to 99.6\%, the highest among all models tested.
GPT-2 Medium and GPT-2 Large exhibit similarly high idle ratios of 98.9\% and 99.1\%, respectively.

\textbf{The parameter server is idle for 79.8\% to 99.6\% of every FL round, depending on model scale.}
For all models beyond ResNet-18, the server spends over 98\% of each round waiting.
A persistent server provisioned for the aggregation workload sits unused during this entire period, yet continues to incur cost.

\begin{tcolorbox}[
  title={Findings for RQ1 -- Part A},
  colback=gray!5,
  colframe=black,
  boxrule=0.4pt,
  arc=2mm,
  left=2mm,right=2mm,top=1mm,bottom=1mm,
  enhanced, breakable
]
79.8\%--99.6\% of every FL round is PS idle time. For VGG-16 and both GPT-2 variants, aggregation accounts for $<$1\% of round time.

\end{tcolorbox}

\subsubsection{Part B: Lambda Deployment Validation}

\mbox{}\\\smallsection{Motivation}
Part~A shows that the parameter server is idle for 79.8\%--99.6\% of each FL round. Serverless platforms such as AWS Lambda are well suited to this workload pattern: functions are instantiated only for the brief aggregation window, billed at 1\,ms granularity, and released immediately after, eliminating the cost of an idle server. However, Part~A measures aggregation on local CPU without network I/O. On Lambda, all data passes through S3, introducing transfer overhead that may change both the time breakdown and the per-round cost. This raises the question of how S3 transfer overhead affects the aggregation time and cost on a real serverless deployment.

\smallsection{Approach}
We deploy the \textsc{GradsSharding} streaming aggregation pipeline on AWS Lambda (us-east-1). For each of the four models, $N=20$ client gradient vectors (random float32) are pre-generated and uploaded to S3. Each Lambda function reads its assigned gradients, performs streaming FedAvg accumulation (one gradient at a time, maintaining a running average), and writes the averaged result back to S3.

Each Lambda function is configured per model:
\begin{itemize}
    \item ResNet-18: 512\,MB memory, $M=1$, 120\,s timeout.
    \item VGG-16: 3{,}008\,MB memory, $M=1$, 900\,s timeout.
    \item GPT-2 Medium: 2{,}048\,MB memory, $M=4$, 900\,s timeout.
    \item GPT-2 Large: 3{,}008\,MB memory, $M=4$, 900\,s timeout.
\end{itemize}
When $M>1$, all shard aggregators are invoked concurrently. Each configuration is executed for 3 rounds with 5 repetitions. The first invocation is excluded as a cold start, yielding 14 warm invocations for statistical reporting.

\begin{figure}[t]
\centering
\includegraphics[width=0.9\columnwidth]{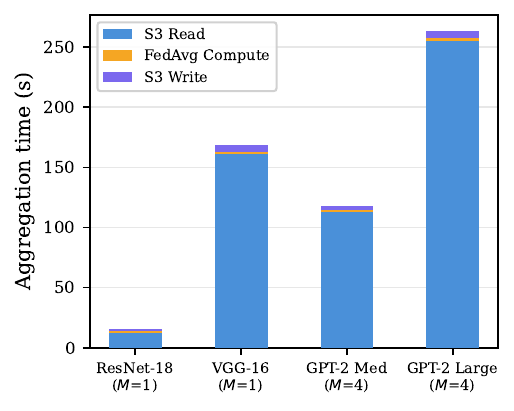}
\caption{Lambda aggregation time breakdown. S3 reads (blue) dominate across all models, accounting for 91--99\% of total time. FedAvg compute (orange) is 1.3--2.9\,s regardless of model size.}
\label{fig:lambda_breakdown}
\end{figure}

Table~\ref{tab:lambda_validation} reports the Lambda deployment results and Figure~\ref{fig:lambda_breakdown} visualizes the time breakdown.

\smallsection{Results}

\textbf{S3 read time accounts for 91\%--99\% of the total Lambda aggregation time across all four models.} FedAvg compute completes in 1.3--2.9\,s regardless of model size, confirming that arithmetic cost is not the bottleneck. The time difference between models is driven by the volume of data transferred: ResNet-18 reads 854\,MB ($20 \times 42.7$\,MB) in 12.6\,s, while GPT-2 Large reads 14.4\,GB ($20 \times 738$\,MB per shard) in 254.9\,s. The effective S3 read throughput ranges from 45 to 68\,MB/s per Lambda function, consistent with single-stream sequential reads within the same region, further confirmed by the VGG-16 shard sweep in Table~\ref{tab:lambda_shard_sweep}.

\begin{table}[h]
\centering
\caption{Lambda aggregation time and cost per model. $N=20$, 14 warm invocations. Cost is Lambda compute only.}
\label{tab:lambda_validation}
\resizebox{\columnwidth}{!}{%
\begin{tabular}{lrrrrrr}
\toprule
\textbf{Model} & $M$ & \shortstack{\textbf{S3 Read}\\\textbf{(s)}} & \shortstack{\textbf{Compute}\\\textbf{(s)}} & \shortstack{\textbf{S3 Write}\\\textbf{(s)}} & \shortstack{\textbf{Total}\\\textbf{(s)}} & \shortstack{\textbf{Cost/1K}\\\textbf{(\$)}} \\
\midrule
ResNet-18    & 1 & 12.6 $\pm$ 1.8 & 1.3 $\pm$ 0.03 & 1.4 $\pm$ 0.2 & 13.9 $\pm$ 1.8  & 0.13 \\
VGG-16       & 1 & 179.9 $\pm$ 21.6 & 2.0 $\pm$ 0.00 & 6.8 $\pm$ 2.5 & 181.9 $\pm$ 21.6 & 8.92 \\
GPT-2 Medium & 4 & 113.0 $\pm$ 13.5 & 1.3 $\pm$ 0.01 & 3.5 $\pm$ 0.3 & 114.3 $\pm$ 13.5 & 15.29 \\
GPT-2 Large  & 4 & 254.9 $\pm$ 26.7 & 2.9 $\pm$ 0.03 & 5.8 $\pm$ 0.4 & 257.8 $\pm$ 26.7 & 50.53 \\
\bottomrule
\end{tabular}%
}
\end{table}

\textbf{Sharding reduces aggregation latency even at fixed S3 throughput.} VGG-16 ($M=1$) takes 181.9\,s because a single function sequentially reads all $20 \times 512$\,MB $\approx$ 10\,GB from S3. GPT-2 Medium has a 2.6$\times$ larger total gradient (1{,}354\,MB), yet with $M=4$ its aggregation completes in 114.3\,s --- 37\% faster than VGG-16, because four concurrent functions each read only $20 \times 338$\,MB. This demonstrates that splitting the gradient across parallel Lambda invocations directly reduces wall-clock time.

\textbf{Lambda compute cost ranges from \$0.13 per 1{,}000 rounds (ResNet-18) to \$50.53 (GPT-2 Large).} Since Lambda billing is proportional to allocated memory $\times$ execution time, and S3 transfer dominates execution time, cost scales with both model size and the number of bytes each function must read. VGG-16 costs \$8.92 per 1{,}000 rounds with $M=1$, while GPT-2 Medium costs \$15.29 with $M=4$: although sharding reduces per-function time, the larger gradient and higher memory allocation (2{,}048\,MB vs.\ 3{,}008\,MB) result in higher total cost.

\begin{tcolorbox}[
  title={Findings for RQ1 -- Part B},
  colback=gray!5,
  colframe=black,
  boxrule=0.4pt,
  arc=2mm,
  left=2mm,right=2mm,top=1mm,bottom=1mm,
  enhanced, breakable
]
91\%--99\% of Lambda aggregation time is S3 I/O. FedAvg compute is 1.3--2.9\,s regardless of model size. S3 read throughput: 45--68\,MB/s per function. Cost per 1{,}000 rounds ranges from \$0.13 (ResNet-18) to \$50.53 (GPT-2 Large), driven by S3 transfer time and memory allocation.
\end{tcolorbox}

\subsection{RQ2: Shard Ablation Study}

\smallsection{Motivation}

RQ1 establishes that serverless aggregation on Lambda is feasible and that S3 transfer, not computation, is the dominant cost factor. RQ1 also shows that sharding reduces wall clock latency: GPT-2 Medium ($M=4$) aggregates 37\% faster than VGG-16 ($M=1$) despite a 2.6$\times$ larger gradient. The shard count $M$ is \textsc{GradsSharding}'s single tunable parameter, controlling a three way tradeoff between per aggregator memory, wall clock latency, and S3 I/O cost. By design, each shard aggregator processes only $|\theta|/M$ parameters, so per aggregator memory is expected to scale as $\mathcal{O}(|\theta|/M)$. This raises two questions: whether the linear memory reduction holds empirically, and how to choose $M$ for a given model size and deployment budget.

\subsubsection{Part A: HPC Ablation}
\mbox{}\\
\smallsection{Approach}
We run \textsc{GradsSharding} with $M \in \{1, 2, 4, 8, 16\}$ shards on HPC hardware, using two models:
\begin{itemize}
    \item ResNet-18 (11.2M params, 43\,MB gradient) trained on CIFAR-100~\cite{krizhevsky2009learning}, non-IID Dirichlet partition ($\alpha = 0.5$), $B=32$.
    \item VGG-16 (134M params, 512\,MB gradient) trained on RVL-CDIP~\cite{harley2015evaluation} document images, $B=8$.
\end{itemize}
Both use $N = 20$ clients for 10 rounds ($\eta = 0.01$, momentum $0.9$, one local epoch). The $M$ shard aggregators execute sequentially on shared hardware. For each configuration we measure:
\begin{itemize}
     \item \textbf{Measured memory:} recorded from the collect-then-average implementation, which loads all $N$ client shards before averaging.
    \item \textbf{Streaming memory:} analytical lower bound assuming a streaming aggregator that holds only two buffers (one incoming shard, one running sum), yielding $2 \text{ buffers} \times (|\theta|/M) \text{ parameters} \times 4 \text{ bytes}$ per aggregator. For example, VGG-16 (512\,MB gradient) at $M=4$: each buffer holds $512/4 = 128$\,MB, so two buffers $= 256$\,MB.
    
    \item Aggregation latency (cumulative, all $M$ aggregators run sequentially).
    \item Estimated Lambda cost (derived from measured time and AWS pricing).
    \item S3 operations per round (analytical): each of $N$ clients uploads $M$ shards ($NM$ PUTs), each of $M$ aggregators reads $N$ shards ($NM$ GETs), each aggregator writes one averaged shard ($M$ PUTs), and each client reads back $M$ averaged shards ($NM$ GETs), totaling $3NM + M$ operations.
\end{itemize}

Table~\ref{tab:rq2_shard} reports the shard ablation results across two model scales. We analyze the effect of the shard count $M$ along three dimensions: memory, latency, and cost.

\smallsection{Results}

\textbf{Memory per aggregator decreases linearly with $M$, confirming the $\mathcal{O}(|\theta|/M)$ bound.} The measured memory column in Table~\ref{tab:rq2_shard} is recorded directly from the collect-then-average implementation. The streaming column is computed analytically from the model's parameter count using the formula defined in the approach.

Both columns halve each time $M$ doubles. For ResNet-18, the streaming bound drops from 85.4\,MB ($M=1$) to 5.3\,MB ($M=16$), fitting within Lambda's 10\,GB limit at any $M$. For VGG-16, streaming memory is 1{,}024\,MB at $M=1$, which fits within the 10\,GB limit, but the collect-then-average implementation requires 10{,}788\,MB at $M=1$, exceeding the ceiling. Sharding to $M=4$ reduces the streaming bound to 256\,MB, leaving ample headroom for implementation overhead and temporary allocations.

\begin{table}[t]
\centering
\caption{Shard ablation for \textsc{GradsSharding}, $N=20$, 10 rounds.}
\label{tab:rq2_shard}
\resizebox{\columnwidth}{!}{%
\begin{tabular}{llrrrrr}
\toprule
\textbf{Model} & \textbf{$M$} & \textbf{\shortstack{Meas.\\Mem (MB)}} & \textbf{\shortstack{Stream.\\Mem (MB)}} & \textbf{\shortstack{Agg.\\Lat.\ (s)}} & \textbf{\shortstack{S3 \\Ops/rnd}} & \textbf{\shortstack{Cost \\(\$)}} \\
\midrule
\multirow{5}{*}{\shortstack{ResNet-18\\(43\,MB)}}
& 1  & 898.8  & 85.4   & 1.15 & 61 & 0.000011 \\
& 2  & 449.4  & 42.7   & 2.35 & 122 & 0.000010 \\
& 4  & 224.7  & 21.4   & 4.63 & 244 & 0.000012 \\
& 8  & 112.4  & 10.7   & 7.61 & 488 & 0.000008 \\
& 16 & 56.2   & 5.3    & 16.65 & 976 & 0.000010 \\
\midrule
\multirow{5}{*}{\shortstack{VGG-16\\(512\,MB)}}
& 1  & 10{,}788 & 1{,}024  & 2.40 & 61 & 0.001055 \\
& 2  & 5{,}394  & 512    & 3.38 & 122 & 0.001004 \\
& 4  & 2{,}697  & 256    & 5.84 & 244 & 0.000712 \\
& 8  & 1{,}349  & 128    & 8.50 & 488 & 0.000347 \\
& 16 & 674      & 64     & 17.16 & 976 & 0.000254 \\
\bottomrule
\end{tabular}%
}
\end{table}

\textbf{Cumulative aggregation latency increases linearly with $M$ under sequential execution: 1.15\,s to 16.65\,s for ResNet-18 and 2.40\,s to 17.16\,s for VGG-16.} Each shard aggregator processes $|\theta|/M$ parameters, but since all $M$ aggregators run sequentially on shared HPC hardware, the total compute time grows with $M$. Dividing the cumulative latency by $M$ gives the average per aggregator time: for VGG-16, $17.16/16 = 1.07$\,s at $M=16$ versus $2.40/1 = 2.40$\,s at $M=1$, showing that each individual aggregator is faster when handling a smaller shard.

\textbf{Lambda compute cost and S3 I/O cost move in opposite directions as $M$ increases, creating a tradeoff that varies with model size.} Lambda billing is proportional to allocated memory $\times$ execution time; as each aggregator handles a smaller shard, both decrease, so Lambda cost drops. For VGG-16, Lambda cost falls from \$0.001055 ($M=1$) to \$0.000254 ($M=16$), a 4.2$\times$ reduction. Conversely, S3 operations grow from 61 to 976 per round ($3NM + M$). 

\begin{figure}[h]
\centering
\includegraphics[width=\columnwidth]{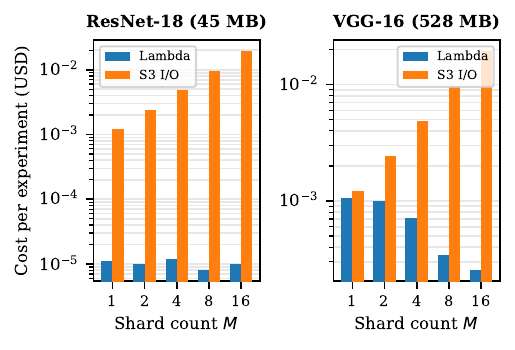}
\caption{Cost breakdown per round: Lambda compute (memory $\times$ time) vs.\ S3 I/O ($3NM + M$ operations), $M \in \{1, 2, 4, 8, 16\}$.}
\label{fig:cost_breakdown}
\end{figure}

Figure~\ref{fig:cost_breakdown} shows that for VGG-16 at $M=1$, Lambda and S3 costs are comparable (\$0.001055 vs.\ \$0.00121); as $M$ grows, Lambda cost drops while S3 cost rises, and by $M=16$ S3 dominates. For ResNet-18, Lambda cost remains below \$0.00001 at all $M$ values, so S3 I/O dominates regardless of shard count. For small models, increasing $M$ adds S3 overhead without reducing Lambda compute cost.

\begin{tcolorbox}[
  title={Findings for RQ2 -- Part A},
  colback=gray!5,
  colframe=black,
  boxrule=0.4pt,
  arc=2mm,
  left=2mm,right=2mm,top=1mm,bottom=1mm,
  enhanced, breakable
]
Memory per aggregator confirms the $\mathcal{O}(|\theta|/M)$ bound. Cumulative latency grows with $M$ sequentially, but per aggregator time decreases. Lambda compute and S3 I/O costs move in opposite directions; for small models, S3 dominates at every $M$.
\end{tcolorbox}

\subsubsection{Part B: Lambda Validation --- Full Shard Sweep}

\mbox{}\\\smallsection{Motivation}\mbox{}\\
Part~A confirms that memory scales as $\mathcal{O}(|\theta|/M)$ and that each individual aggregator is faster with fewer parameters, but cumulative latency grows with $M$ because aggregators run sequentially on shared HPC hardware. On AWS Lambda, each aggregator executes as an independent function, so all $M$ aggregators can run concurrently. Part~A also estimates Lambda and S3 costs analytically; deploying on real Lambda infrastructure validates whether these estimates hold under actual S3 throughput and concurrent execution.

\smallsection{Approach}\mbox{}\\
We deploy VGG-16 (134M params, 512\,MB gradient) on AWS Lambda with $M \in \{1, 2, 4, 8, 16\}$ shards and $N=20$ clients. Each shard aggregator runs as an independent Lambda function (3{,}008\,MB memory, 900\,s timeout) in us-east-1. For $M > 1$, all $M$ aggregators are invoked concurrently. We run 3 rounds with 5 repetitions per configuration; the first invocation is excluded to isolate cold start effects, yielding 14 warm invocations per $M$.

\textit{Collected metrics.}
For each $M$ we report:
\begin{itemize}
    \item \textbf{Time breakdown:} S3 read time, FedAvg compute time, and S3 write time per aggregator.
    \item \textbf{Speedup:} wall clock aggregation time relative to $M=1$.
    \item \textbf{S3 operations per round:} $3NM + M$ (as derived in Part~A).
    \item \textbf{Cost per 1{,}000 rounds:} Lambda compute (memory $\times$ time) + S3 I/O, using actual AWS billing.
\end{itemize}

\begin{figure}[t]
\centering
\includegraphics[width=\columnwidth]{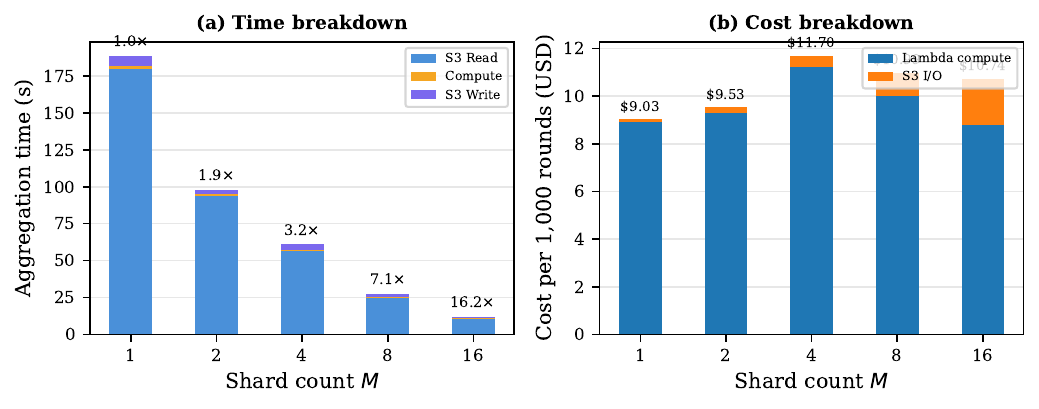}
\caption{VGG-16 shard sweep on AWS Lambda ($N=20$). (a)~Aggregation time decomposed into S3 read, FedAvg compute, and S3 write; speedup annotations relative to $M=1$. (b)~Cost per 1{,}000 rounds decomposed into Lambda compute and S3 I/O.}
\label{fig:lambda_vgg16_shard_sweep}
\end{figure}

Table~\ref{tab:lambda_shard_sweep} reports the full sweep results and Figure~\ref{fig:lambda_vgg16_shard_sweep} visualizes the time and cost breakdown. Three findings emerge from this deployment.

\mbox{}\\\smallsection{Results}

\textbf{Aggregation time decreases from 181.9\,s at $M=1$ to 11.2\,s at $M=16$, a 16.2$\times$ speedup.} This confirms that the cumulative latency increase observed in Part~A is an artifact of sequential execution on shared hardware. On Lambda, each shard aggregator runs as an independent concurrent function, so wall clock time is determined by the slowest aggregator rather than the sum. The speedup scales near-linearly with $M$: 1.9$\times$ at $M=2$, 3.2$\times$ at $M=4$, 7.1$\times$ at $M=8$, and 16.2$\times$ at $M=16$.

\begin{figure}[t]
\centering
\includegraphics[width=\columnwidth]{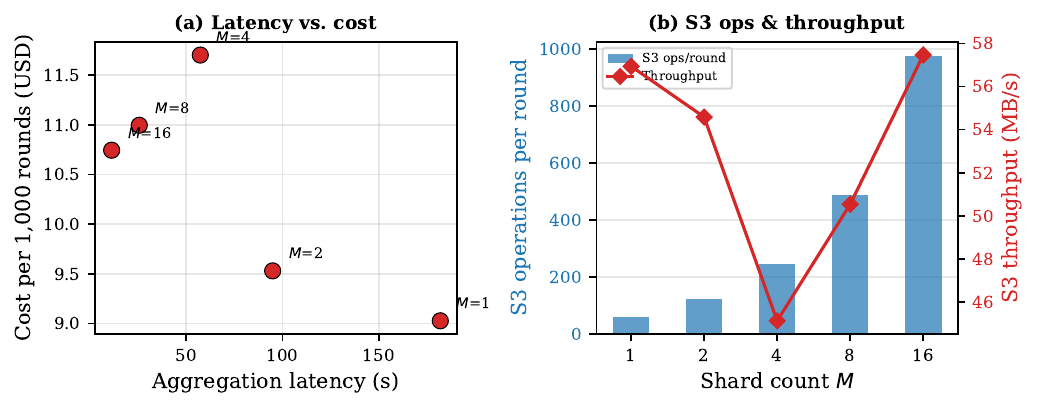}
\caption{(a)~Latency vs.\ cost tradeoff: $M=4$ is the least cost-efficient; $M=1$ and $M=16$ are Pareto-optimal for cost and latency, respectively. (b)~S3 operations per round grow linearly with $M$, while per-function throughput remains stable at 45--58\,MB/s.}
\label{fig:lambda_vgg16_tradeoff}
\end{figure}

\textbf{S3 read time accounts for 98.9--99.1\% of total aggregation time at every shard count.} FedAvg compute scales linearly with shard size (1.96\,s at $M=1$ to 0.13\,s at $M=16$) but remains below 2\,s throughout. The effective per-function S3 throughput ranges from 45.1 to 57.5\,MB/s across configurations (Figure~\ref{fig:lambda_vgg16_tradeoff}b), as each function reads its shard objects one at a time from S3. This uniform dominance means that S3 throughput, not compute capacity, is the binding constraint for serverless aggregation regardless of the shard count.

\begin{table}[h]
\centering
\caption{Lambda shard sweep for VGG-16 (512\,MB gradient), $N=20$. Latency speedup relative to $M=1$.}
\label{tab:lambda_shard_sweep}
\resizebox{\columnwidth}{!}{%
\begin{tabular}{rrrrrrrr}
\toprule
$M$ & \textbf{\shortstack{Shard\\(MB)}} & \textbf{\shortstack{S3 Read\\(s)}} & \textbf{\shortstack{Compute\\ms (s)}} & \textbf{\shortstack{S3 Write\\(s)}} & \textbf{\shortstack{Latency\\Speedup}} & \textbf{\shortstack{S3\\Ops}} & \textbf{\shortstack{Cost/1K\\(\$)}} \\
\midrule
1  & 512.3 & 179.9 $\pm$ 21.6 & 1963 (1.96) & 6.8 $\pm$ 2.5 & 1.0$\times$ & 61  & 9.03 \\
2  & 256.2 & 93.9 $\pm$ 12.2  & 1000 (1.00) & 3.0 $\pm$ 0.4 & 1.9$\times$ & 122 & 9.53 \\
4  & 128.1 & 56.8 $\pm$ 8.4   & 510 (0.51)  & 3.7 $\pm$ 2.0 & 3.2$\times$ & 244 & 11.70 \\
8  & 64.0  & 25.3 $\pm$ 4.3   & 260 (0.26)  & 2.0 $\pm$ 1.6 & 7.1$\times$ & 488 & 11.00 \\
16 & 32.0  & 11.1 $\pm$ 1.4   & 132 (0.13)  & 0.7 $\pm$ 0.2 & 16.2$\times$ & 976 & 10.74 \\
\bottomrule
\end{tabular}%
}
\end{table}

\textbf{The cost per 1{,}000 rounds peaks at $M=4$ (\$11.70) and decreases toward both ends: $M=1$ is cheapest at \$9.03, and $M=16$ drops back to \$10.74} (Figure~\ref{fig:lambda_vgg16_shard_sweep}b). This shape arises because Lambda compute cost is proportional to $M \times \text{memory} \times \text{time}$: increasing $M$ reduces both per-function memory and per-function execution time, but the number of concurrent functions increases. At intermediate $M$, the $M$ multiplier dominates; at high $M$, each function becomes so small and fast that Lambda compute drops back down. Meanwhile, S3 I/O cost grows linearly from \$0.12 at $M=1$ to \$1.94 at $M=16$, rising from 1.3\% to 18\% of total cost. Figure~\ref{fig:lambda_vgg16_tradeoff}a visualizes this tradeoff: $M=1$ and $M=16$ are Pareto-optimal for cost and latency, respectively, while $M=4$ is the least cost-efficient. Practitioners optimizing for \emph{latency} should prefer $M=16$ (16.2$\times$ faster at a 19\% cost premium over $M=1$), while those optimizing for \emph{cost} should prefer $M=1$ when latency constraints allow a 3-minute aggregation window.

\begin{tcolorbox}[
  title={Findings for RQ2 -- Part B},
  colback=gray!5,
  colframe=black,
  boxrule=0.4pt,
  arc=2mm,
  left=2mm,right=2mm,top=1mm,bottom=1mm,
  enhanced, breakable
]
Concurrent execution on Lambda yields 16.2$\times$ latency speedup at $M=16$ with 19\% cost premium over $M=1$. S3 read time accounts for 98.9--99.1\% of aggregation time at every $M$. Per-function S3 throughput: 45--58\,MB/s. Cost per 1{,}000 rounds: \$9.03 at $M=1$ (cheapest), \$10.74 at $M=16$ (fastest).
\end{tcolorbox}

\subsection{RQ3: Cross-Architecture Cost \& Scalability}

\mbox{}\\\smallsection{Motivation}\mbox{}\\
RQ1 and RQ2 evaluate \textsc{GradsSharding} in isolation: memory per aggregator scales as $\mathcal{O}(|\theta|/M)$, concurrent Lambda execution yields near-linear speedup with $M$, and S3 I/O dominates aggregation time. These results characterize \textsc{GradsSharding}'s own scaling behavior but do not reveal how it compares to existing serverless FL architectures on real cloud infrastructure. Two competing approaches have been proposed: $\lambda$-FL~\cite{jayaram2022lambda}, which uses a two-level tree with $\lceil\sqrt{N}\rceil$ leaf aggregators plus a root, and LIFL~\cite{ciobotaru2024lifl}, which extends this to a three-level hierarchy with $\lceil\sqrt[3]{N}\rceil$ branching at each level. Both architectures require each aggregator to hold the \emph{full} gradient in memory, while \textsc{GradsSharding} partitions the gradient across $M$ independent aggregators. As model sizes grow, this difference in memory footprint may determine which architectures remain deployable within Lambda's 10{,}240\,MB (10\,GB) memory limit. Comparing all three on real AWS Lambda infrastructure, across a range of model sizes, reveals whether \textsc{GradsSharding}'s memory advantage translates into concrete cost and feasibility benefits.

\smallsection{Approach}\mbox{}\\
We deploy all three architectures on AWS Lambda (us-east-1) using a unified codebase and the same streaming aggregation logic: each aggregator reads one client gradient at a time from S3, accumulates a running sum, and divides at the end. During streaming, two persistent buffers coexist in memory: the running sum and the incoming client gradient being read from S3. However, transient allocations during S3 response parsing and NumPy deserialization create additional memory pressure, so in practice each aggregator requires approximately $3 \times \text{input\_size} + \text{overhead}$ bytes, where the overhead ($\sim$450\,MB) accounts for the AWS Lambda runtime and the AWSSDKPandas layer.

We test four model sizes spanning two orders of magnitude in gradient size, with $N=20$ clients:
\begin{itemize}
    \item \textbf{ResNet-18} (11.2M params, 42.7\,MB gradient)
    \item \textbf{VGG-16} (134M params, 512.3\,MB gradient)
    \item \textbf{GPT-2 Large} (774M params, 2{,}953\,MB gradient)
    \item \textbf{Synthetic 5\,GB} (1.34B params, 5{,}120\,MB gradient)
\end{itemize}

\textit{Architecture deployment.} For each model, we measure the complete round-trip cost of one FL aggregation round, including client uploads, Lambda aggregation, and client read-back. The three architectures differ in topology, S3 access pattern, and memory footprint per aggregator (memory: $3 \times \text{input\_size} + 450$\,MB). Below, we describe the steps and S3 operations for each.

\textbf{$\lambda$-FL} ($k = \lceil\sqrt{N}\rceil = 5$, yielding 4 leaves + 1 root):
\begin{enumerate}
    \item $N$ clients upload their full gradients to S3 $\rightarrow$ $N$ PUTs.
    \item 4~leaf Lambda functions are invoked concurrently; each reads 5~client gradients from S3, computes a partial average, and writes the result to S3 $\rightarrow$ $N$~GETs + 4~PUTs.
    \item 1~root Lambda reads the 4~leaf results from S3, averages them, and writes the final global gradient $\rightarrow$ 4~GETs + 1~PUT.
    \item $N$~clients read the final averaged gradient from S3 $\rightarrow$ $N$~GETs.
\end{enumerate}
\textit{Total:} 25~PUTs + 44~GETs = 69 S3 operations. 5~Lambda invocations across 2 sequential phases. Memory per function: $3 \times |\theta| \times 4\text{\,bytes} + 450\text{\,MB}$.

\textbf{LIFL} (branching factor $\lceil\sqrt[3]{N}\rceil = 3$, yielding 7~L1 + 3~L2 + 1~root):
\begin{enumerate}
    \item $N$~clients upload their full gradients to S3 $\rightarrow$ $N$~PUTs.
    \item 7~level-1 Lambda functions each read 2--3~client gradients, average, and write to S3 $\rightarrow$ $N$~GETs + 7~PUTs.
    \item 3~level-2 Lambda functions each read 2--3~level-1 results, average, and write to S3 $\rightarrow$ 7~GETs + 3~PUTs.
    \item 1~root Lambda reads the 3~level-2 results, averages, and writes the final gradient $\rightarrow$ 3~GETs + 1~PUT.
    \item $N$~clients read the final averaged gradient from S3 $\rightarrow$ $N$~GETs.
\end{enumerate}
\textit{Total:} 31~PUTs + 50~GETs = 81 S3 operations. 11~Lambda invocations across 3 sequential phases. Memory per function: same as $\lambda$-FL.

\textbf{\textsc{GradsSharding}} ($M=4$ for ResNet-18/VGG-16/GPT-2 Large, $M=8$ for Synthetic~5\,GB):
\begin{enumerate}
    \item $N$~clients each split their gradient into $M$ shards and upload all shards to S3 $\rightarrow$ $NM$~PUTs.
    \item $M$~Lambda functions are invoked \emph{concurrently}; each reads its shard index from all $N$~clients, accumulates a running average, and writes the averaged shard to S3 $\rightarrow$ $NM$~GETs + $M$~PUTs.
    \item $N$~clients read back all $M$~averaged shards from S3 and concatenate them to reconstruct the full aggregated gradient $\rightarrow$ $NM$~GETs.
\end{enumerate}
\textit{Total ($M\!=\!4$):} 84~PUTs + 160~GETs = 244 S3 operations. 4~Lambda invocations, all concurrent (single phase). Memory per function: $3 \times (|\theta|/M) \times 4\text{\,bytes} + 450\text{\,MB}$.

\textit{Feasibility.} Before deployment, we compute the peak memory for each architecture--model pair using the empirically validated formula $3 \times \text{input\_size} + 450$\,MB. If peak memory exceeds Lambda's 10{,}240\,MB maximum, the configuration is \emph{infeasible}. For $\lambda$-FL and LIFL, each aggregator processes the full gradient: Synthetic~5\,GB requires $3 \times 5{,}120 + 450 = 15{,}810$\,MB, exceeding the platform limit. GPT-2 Large requires $3 \times 2{,}953 + 450 = 9{,}309$\,MB---within the 10{,}240\,MB maximum but consuming 91\% of available memory. We deploy only \textsc{GradsSharding} at these scales: with $M=4$, each GPT-2 Large aggregator requires only $3 \times 738 + 450 = 2{,}665$\,MB, and with $M=8$, each Synthetic~5\,GB aggregator requires 2{,}370\,MB.

\textit{Metrics.} For each feasible configuration, we run 3 rounds $\times$ 3 repetitions. The first invocation is excluded as a cold start, yielding 8 warm measurements. We record:
\begin{itemize}
    \item \textbf{Wall-clock time:} end-to-end aggregation latency (from first Lambda invocation to last result written to S3).
    \item \textbf{Time breakdown:} S3 read time, FedAvg compute time, and S3 write time per aggregator.
    \item \textbf{Cost per round:} Lambda compute (allocated memory $\times$ billed duration $\times$ \$0.0000166667/GB-s) + \emph{full round-trip} S3 I/O including client uploads, aggregator reads/writes, and client read-back (\$0.005/1K PUTs, \$0.0004/1K GETs).
\end{itemize}

\begin{figure}[t]
\centering
\includegraphics[width=\columnwidth]{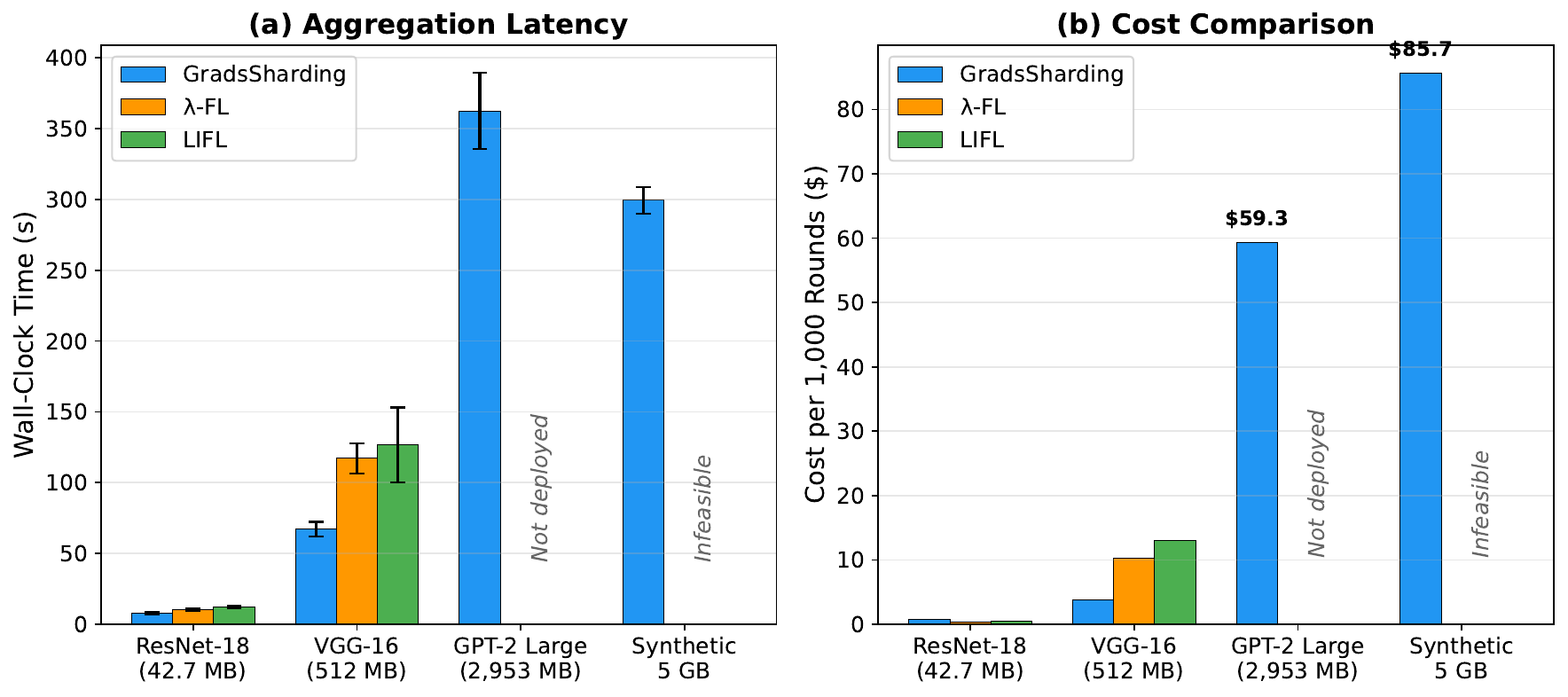}
\caption{Cross-architecture comparison on AWS Lambda ($N=20$). (a)~Aggregation wall-clock time and (b)~cost per 1{,}000 rounds across model sizes. $\lambda$-FL and LIFL were not deployed for GPT-2 Large (91\% of Lambda's memory) or Synthetic 5\,GB (exceeds Lambda's 10\,GB limit).}
\label{fig:rq3_comparison}
\end{figure}

Table~\ref{tab:rq3_lambda} reports the results across all four model sizes. Figure~\ref{fig:rq3_comparison} visualizes the latency and cost comparison.
\mbox{}\\
\smallsection{Results}

\begin{table*}[t]
\centering
\caption{Cross-architecture comparison on AWS Lambda, $N=20$ clients, 8 warm measurements. ``---'' indicates the architecture was not deployed at this scale (see footnotes). Cost includes Lambda compute + full round-trip S3 I/O (client uploads, aggregator reads/writes, and client read-back).}
\label{tab:rq3_lambda}
\resizebox{\textwidth}{!}{%
\begin{tabular}{llrrrrrrr}
\toprule
\textbf{Model} & \textbf{Architecture} & \shortstack{\textbf{Memory}\\\textbf{(MB)}} & \shortstack{\textbf{\# Lambda}\\\textbf{Invoc.}} & \shortstack{\textbf{S3 Ops}\\\textbf{(PUTs/GETs)}} & \shortstack{\textbf{Wall Clock}\\\textbf{(s)}} & \shortstack{\textbf{Cost/Round}\\\textbf{(\$)}} & \shortstack{\textbf{Cost/1K}\\\textbf{Rounds (\$)}} & \textbf{Feasible} \\
\midrule
\multirow{3}{*}{\shortstack[l]{ResNet-18\\(42.7\,MB)}}
& \textsc{GradsSharding} ($M\!=\!4$) & 472   & 4  & 84/160  & \textbf{7.9 $\pm$ 0.8}   & 0.000695 & 0.70 & \checkmark \\
& $\lambda$-FL                        & 536   & 5  & 25/44   & 10.4 $\pm$ 0.8  & \textbf{0.000378} & \textbf{0.38} & \checkmark \\
& LIFL                                 & 536   & 11 & 31/50   & 12.3 $\pm$ 1.1  & 0.000523 & 0.52 & \checkmark \\
\midrule
\multirow{3}{*}{\shortstack[l]{VGG-16\\(512.3\,MB)}}
& \textsc{GradsSharding} ($M\!=\!4$) & \textbf{835}   & 4  & 84/160  & \textbf{67.2 $\pm$ 5.1}  & \textbf{0.00382}  & \textbf{3.82} & \checkmark \\
& $\lambda$-FL                        & 1{,}987 & 5  & 25/44   & 117.3 $\pm$ 10.7 & 0.01028  & 10.28 & \checkmark \\
& LIFL                                 & 1{,}987 & 11 & 31/50   & 126.6 $\pm$ 26.5 & 0.01303  & 13.03 & \checkmark \\
\midrule
\multirow{3}{*}{\shortstack[l]{GPT-2 Large\\(2{,}953\,MB)}}
& \textsc{GradsSharding} ($M\!=\!4$) & \textbf{2{,}665} & 4  & 84/160  & \textbf{362.5 $\pm$ 26.8} & \textbf{0.05929}  & \textbf{59.29} & \checkmark \\
& $\lambda$-FL                        & ---   & --- & ---     & ---              & ---      & ---   & $\times$\textsuperscript{a} \\
& LIFL                                 & ---   & --- & ---     & ---              & ---      & ---   & $\times$\textsuperscript{a} \\
\midrule
\multirow{3}{*}{\shortstack[l]{Synthetic\\5\,GB (5{,}120\,MB)}}
& \textsc{GradsSharding} ($M\!=\!8$) & \textbf{2{,}370} & 8  & 168/320 & \textbf{299.4 $\pm$ 9.5}  & \textbf{0.08566}  & \textbf{85.66} & \checkmark \\
& $\lambda$-FL                        & ---   & --- & ---     & ---              & ---      & ---   & $\times$\textsuperscript{b} \\
& LIFL                                 & ---   & --- & ---     & ---              & ---      & ---   & $\times$\textsuperscript{b} \\
\bottomrule
\multicolumn{8}{l}{\footnotesize \textsuperscript{a} Requires 9{,}309\,MB per aggregator ($3 \times 2{,}953 + 450$); within Lambda's 10\,GB max but at 91\% utilization. Not deployed.} \\
\multicolumn{8}{l}{\footnotesize \textsuperscript{b} Requires 15{,}810\,MB per aggregator ($3 \times 5{,}120 + 450$), exceeds Lambda's 10{,}240\,MB maximum.}
\end{tabular}%
}
\end{table*}

\textbf{At small model sizes, all three architectures are feasible; $\lambda$-FL is cheapest due to fewer S3 operations.} For ResNet-18 (42.7\,MB), all architectures fit comfortably within Lambda's memory limit. $\lambda$-FL achieves the lowest cost at \$0.38 per 1{,}000 rounds, followed by LIFL (\$0.52) and \textsc{GradsSharding} (\$0.70). At this scale, Lambda compute is small across all configurations (all allocate under 536\,MB), so the cost is dominated by S3 I/O. \textsc{GradsSharding}'s higher cost stems from its 244 S3 operations per round (84~PUTs + 160~GETs) versus 69 for $\lambda$-FL and 81 for LIFL, because each of $N$~clients uploads $M$~shards and reads back $M$~averaged shards. Despite higher cost, \textsc{GradsSharding} achieves the lowest latency (7.9\,s vs.\ 10.4\,s for $\lambda$-FL and 12.3\,s for LIFL) because its single-phase concurrent execution avoids the sequential dependencies of tree-based approaches.

\textbf{At medium model sizes, \textsc{GradsSharding} is 1.7$\times$ faster and 2.7$\times$ cheaper than $\lambda$-FL.} For VGG-16 (512.3\,MB), \textsc{GradsSharding} aggregates in 67.2\,s at \$3.82 per 1{,}000 rounds, while $\lambda$-FL requires 117.3\,s at \$10.28 and LIFL takes 126.6\,s at \$13.03. The cost gap is now driven by per-function memory allocation: each \textsc{GradsSharding} aggregator allocates 835\,MB (for a 128\,MB shard), while $\lambda$-FL and LIFL allocate 1{,}987\,MB each (for the full 512\,MB gradient). Since Lambda billing scales with memory $\times$ time, the 2.4$\times$ reduction in per-function memory compounds with the 1.7$\times$ shorter execution time, and the Lambda compute savings far outweigh \textsc{GradsSharding}'s higher S3 I/O cost (244 vs.\ 69--81 operations).

$\lambda$-FL's two-phase execution (leaves then root) also contributes to its higher latency: the root aggregator cannot begin until all leaves complete, adding a sequential dependency that does not exist in \textsc{GradsSharding}'s flat-parallel model. LIFL's three-phase hierarchy amplifies this effect further, and its high variance ($\pm$26.5\,s) reflects the sensitivity of multi-level pipelines to straggler aggregators at any level.

\textbf{At large model sizes, \textsc{GradsSharding}'s memory advantage becomes decisive.} For GPT-2 Large (2{,}953\,MB gradient), each $\lambda$-FL or LIFL aggregator would require $3 \times 2{,}953 + 450 = 9{,}309$\,MB of peak memory during streaming accumulation---91\% of Lambda's 10{,}240\,MB maximum, leaving minimal headroom for runtime variability. \textsc{GradsSharding} with $M=4$ requires only 2{,}665\,MB per aggregator, a 3.5$\times$ reduction in per-function memory that translates directly to proportional cost savings under Lambda's memory~$\times$~time billing. We deploy only \textsc{GradsSharding} at this scale, which completes aggregation in 362.5\,s at \$59.29 per 1{,}000 rounds.

\textbf{Beyond $\sim$3\,GB gradient size, $\lambda$-FL and LIFL exceed Lambda's 10\,GB memory limit entirely.} For the Synthetic 5\,GB workload, each full-gradient aggregator would require $3 \times 5{,}120 + 450 = 15{,}810$\,MB, more than 1.5$\times$ the platform maximum. No amount of tree restructuring resolves this: the constraint is inherent to any architecture where a single aggregator must process the full gradient. \textsc{GradsSharding} with $M=8$ splits the gradient into 640\,MB shards, each requiring 2{,}370\,MB, and aggregates in 299.4\,s at \$85.66 per 1{,}000 rounds. The lower wall-clock time compared to GPT-2 Large (299\,s vs.\ 363\,s) despite the larger total data volume is explained by the higher degree of parallelism ($M=8$ vs.\ $M=4$): eight concurrent functions each read less data than four functions processing larger shards.

\textbf{The feasibility boundary is determined by the relationship between gradient size and Lambda's memory limit.} Using the empirical $3\times$ memory formula, the maximum gradient size that a full-gradient architecture can handle is $(10{,}240 - 450) / 3 \approx 3{,}263$\,MB. Any model with a gradient exceeding this threshold cannot be aggregated by $\lambda$-FL or LIFL on Lambda, regardless of the number of clients or tree topology. Below this hard limit, large gradients like GPT-2 Large (2{,}953\,MB) still consume over 90\% of Lambda's maximum memory, making full-gradient architectures impractical at scale. \textsc{GradsSharding} circumvents this limit entirely: by choosing $M$ such that each shard fits comfortably within the memory budget, it can aggregate arbitrarily large models. For a 10\,GB gradient, $M=16$ yields 640\,MB shards; for a 100\,GB gradient, $M=128$ yields 800\,MB shards, both well within Lambda's capacity.

\begin{tcolorbox}[
  title={Findings for RQ3},
  colback=gray!5,
  colframe=black,
  boxrule=0.4pt,
  arc=2mm,
  left=2mm,right=2mm,top=1mm,bottom=1mm,
  enhanced, breakable
]
On real AWS Lambda with $N=20$ clients and full round-trip S3 costs: at ResNet-18 scale, $\lambda$-FL is cheapest (\$0.38/1K vs.\ \$0.70/1K) due to fewer S3 operations. At VGG-16 scale, \textsc{GradsSharding} is 1.7$\times$ faster and 2.7$\times$ cheaper than $\lambda$-FL (\$3.82/1K vs.\ \$10.28/1K), as Lambda compute savings outweigh the S3 I/O overhead. At GPT-2 Large scale, $\lambda$-FL and LIFL would require 9{,}309\,MB per aggregator (91\% of Lambda's 10\,GB maximum) versus 2{,}665\,MB for \textsc{GradsSharding}. Beyond $\sim$3\,GB gradient size, full-gradient architectures exceed Lambda's 10\,GB limit entirely. \textsc{GradsSharding} remains deployable at all scales by increasing $M$, with no architectural ceiling on model size.
\end{tcolorbox}

\section{Discussion}
\label{sec:discussion}

\subsection{Practical Architecture Selection}

Our results across all three RQs yield a concrete decision framework for practitioners deploying serverless FL aggregation.

For \textbf{small models} (gradient $<$500\,MB), any architecture works. All three fit within Lambda's memory limit, and the cost differences are marginal (under \$1 per 1{,}000 rounds). $\lambda$-FL is the cheapest option at this scale (\$0.38/1K for ResNet-18) because it issues far fewer S3 operations (69 vs.\ 244 for \textsc{GradsSharding}). However, \textsc{GradsSharding} offers the lowest latency (7.9\,s vs.\ 10.4\,s) due to its single-phase concurrent execution, which may matter for latency-sensitive deployments.

For \textbf{medium models} (500\,MB--3\,GB gradient), \textsc{GradsSharding} dominates on both cost and latency. At VGG-16 scale, it is 2.7$\times$ cheaper and 1.7$\times$ faster than $\lambda$-FL. The crossover is driven by a fundamental asymmetry: S3 operations have a fixed per-request price regardless of object size, while Lambda compute scales with memory $\times$ time. As model size grows, Lambda compute dominates total cost, and \textsc{GradsSharding}'s $M$-fold memory reduction per aggregator translates directly into savings that outweigh the higher S3 operation count.

For \textbf{large models} (gradient $>$3\,GB), \textsc{GradsSharding} is the only viable option. The empirical memory formula $3 \times \text{input\_size} + 450$\,MB yields a hard feasibility threshold of approximately $(10{,}240 - 450) / 3 \approx 3{,}263$\,MB for full-gradient architectures on Lambda's 10\,GB maximum. Even below this hard limit, full-gradient architectures become impractical: at GPT-2 Large (2{,}953\,MB), $\lambda$-FL and LIFL require 9{,}309\,MB per aggregator---91\% of Lambda's maximum, with 3.5$\times$ higher per-function memory cost than \textsc{GradsSharding}'s 2{,}665\,MB ($M=4$). Beyond the $\sim$3{,}263\,MB threshold, full-gradient architectures exceed Lambda's memory limit entirely, regardless of tree topology.

\subsection{Choosing the Shard Count $M$}

The RQ2 shard sweep and RQ3 cross-architecture comparison together provide practical guidance for selecting $M$. Two constraints must be satisfied simultaneously: (1)~per-aggregator memory must fit within the Lambda limit, and (2)~total cost should be minimized given a latency requirement.

For \textbf{latency}, higher $M$ is always faster. The Lambda sweep confirms near-linear scaling from 181.9\,s at $M=1$ to 11.2\,s at $M=16$ (16.2$\times$ speedup) for VGG-16. Since all $M$ aggregators run concurrently, wall-clock aggregation time decreases monotonically with $M$. The RQ3 results reinforce this: the Synthetic 5\,GB model aggregates in 299\,s with $M=8$ versus 363\,s for GPT-2 Large with $M=4$, despite having 1.7$\times$ more data, because eight concurrent functions each process smaller shards.

For \textbf{cost}, the picture is more nuanced. Lambda bills $M \times \text{memory} \times \text{time}$ per round; increasing $M$ reduces both per-function memory and time, but the $M$ multiplier dominates at intermediate shard counts. For VGG-16, cost peaks at $M=4$ (\$11.70/1K) and decreases toward both ends: $M=1$ is cheapest at \$9.03/1K, while $M=16$ drops to \$10.74/1K. Meanwhile, S3 I/O cost grows linearly with $M$ (from \$0.12 at $M=1$ to \$1.94 at $M=16$). Practitioners optimizing for latency should prefer high $M$ (16.2$\times$ faster at a 19\% cost premium), while those tolerating a 3-minute aggregation window should use $M=1$.

\subsection{S3 as the Bottleneck}

Across all experiments, S3 I/O is the dominant cost and latency factor. In the RQ1 Lambda deployment, S3 reads account for 91--99\% of total execution time. In the RQ2 shard sweep, this fraction holds at 98.9--99.1\% regardless of $M$. In the RQ3 cross-architecture comparison, the same pattern persists: FedAvg compute completes in under 3\,s across all model sizes, while S3 transfers take tens to hundreds of seconds.

The measured single-stream S3 read throughput of 45--58\,MB/s per Lambda function is consistent across all experiments and shard counts (Figure~\ref{fig:lambda_vgg16_tradeoff}b), indicating that concurrent Lambda functions do not contend for S3 bandwidth. This has two implications.

First, \textbf{optimizing compute is irrelevant}. Using SIMD-optimized kernels or GPU-accelerated Lambda would yield marginal improvements, since aggregation arithmetic is already $<$1\% of execution time. Effort should instead focus on parallelizing S3 reads within each aggregator (byte-range requests, multi-part downloads) or batching multiple client gradients into a single S3 object.

Second, \textbf{bypassing S3} for small payloads is promising. Lambda supports up to 6\,MB synchronous invocation payloads. For small models or heavily sharded large models, clients could send gradient shards directly as invocation payloads, eliminating S3 latency and cost entirely. A hybrid approach---direct payload for shards under 6\,MB, S3 for larger ones---could substantially improve the cost-latency tradeoff, particularly at high $M$ where individual shards become small.

\subsection{Cold Start Overhead}

The Lambda deployment reveals that cold start overhead is modest relative to total aggregation time, adding 2--4\,s on top of S3 transfer time for the first invocation. Because S3 I/O dominates, this represents a small fraction of the total round. In a production FL deployment, cold starts affect only the first round after a long idle period. Since training typically runs for hundreds of consecutive rounds, the cost is well amortized. For latency-sensitive deployments, Lambda's provisioned concurrency feature can eliminate cold starts entirely: the VGG-16 $M=4$ configuration would require only 4 provisioned instances.

\section{Threats to Validity}
\label{sec:threats}

\smallsection{Reimplementation of baselines}\mbox{}\\ We reimplemented $\lambda$-FL and LIFL from their published descriptions since neither codebase is publicly available. While we followed the published designs closely, our implementations may differ from the originals in low-level optimizations. To mitigate this risk, all three systems share a common codebase for everything except the aggregation topology (Section~\ref{sec:setup}), and the RQ3 Lambda deployment uses identical streaming aggregation logic across all architectures. LIFL's shared-memory optimization requires colocation of aggregator functions on the same physical node, which is unavailable on Lambda where each function runs in an isolated container. Our Lambda deployment adapts LIFL by replacing shared-memory transfers with S3 operations.

\smallsection{Memory formula}\mbox{}\\ The empirical $3\times$ memory formula used for feasibility analysis was derived from observed OOM failures and may not generalize exactly to other Lambda runtimes or library versions. The 450\,MB overhead accounts for the Python 3.12 runtime with the AWSSDKPandas layer; different configurations may require recalibration. However, the formula is conservative (no OOM failures occurred at the computed memory levels), and the qualitative conclusion---that full-gradient architectures hit Lambda's memory ceiling at multi-gigabyte scale---holds regardless of the exact multiplier.

\smallsection{Sequential client training} In our HPC experiments (RQ1 Part~A, RQ2 Part~A), clients train sequentially on a shared GPU, whereas in a distributed deployment clients would train in parallel. This inflates measured round times but does not affect the aggregation-phase measurements, which are the primary focus of comparison.

\smallsection{Model scope}\mbox{}\\ We evaluate four model sizes spanning two orders of magnitude (43\,MB to 5\,GB) on real AWS Lambda, covering both CNN (ResNet-18, VGG-16) and transformer-scale (GPT-2 Large) gradient sizes. Models with heterogeneous layer sizes (e.g., mixture-of-experts) may produce imbalanced shards under uniform gradient partitioning. Adaptive sharding strategies are left for future work.

\smallsection{Client count}\mbox{}\\ All RQ3 experiments use $N=20$ clients. Increasing $N$ raises S3 operation counts proportionally for all architectures and increases per-aggregator data volume, which may shift the cost crossover point. However, the memory feasibility boundary depends on gradient size and $M$, not $N$, so the scalability wall conclusion is unaffected.

\section{Conclusion}
\label{sec:conclusion}

We presented \textsc{GradsSharding}, a serverless FL aggregation strategy that partitions the gradient tensor across $M$ parallel Lambda functions and bounds per-function memory at $\mathcal{O}(|\theta|/M)$, independent of client count. Because FedAvg averaging is element-wise, the sharded and tree-based topologies produce bit-identical aggregated gradients, so the comparison is purely a systems-level evaluation.

Three findings emerge from our deployment on real AWS Lambda. First, the parameter server is idle for 80--99.6\% of every FL round (RQ1), confirming the serverless premise: provisioning a persistent server for a workload that is active less than 1\% of the time is wasteful. Second, \textsc{GradsSharding}'s concurrent shard execution delivers near-linear latency scaling with $M$ (16.2$\times$ speedup at $M=16$), with S3 I/O accounting for over 99\% of aggregation time and FedAvg compute completing in under 3\,s regardless of model size (RQ2). Third, the cross-architecture comparison (RQ3) reveals a cost crossover: below $\sim$500\,MB, $\lambda$-FL is cheapest due to fewer S3 operations; above it, \textsc{GradsSharding}'s per-function memory savings dominate, reaching 2.7$\times$ cost reduction at VGG-16 scale. Beyond $\sim$3\,GB gradient size, $\lambda$-FL and LIFL cannot fit a streaming aggregator within Lambda's memory limit, while \textsc{GradsSharding} continues to operate by increasing $M$.

Future work includes adaptive shard counts that respond to model and memory conditions at runtime, composition with gradient compression to reduce S3 transfer volume, optimizing S3 access patterns (parallel byte-range reads, direct Lambda invocation payloads for small shards) to reduce the I/O bottleneck, and non-uniform sharding for models with heterogeneous layer sizes.

\section*{Acknowledgment}
This work was partially supported by a Faculty Research Fellowship from the University Research Committee (URC) at Oakland University.

\bibliographystyle{ieeetr}

\bibliography{references}

\begin{IEEEbiography}[{\includegraphics[width=1in,height=1.25in,clip,keepaspectratio]{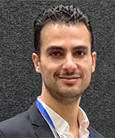}}]{Amine Barrak}
is an Assistant Professor with the Department of Computer Science and Engineering, Oakland University, Rochester, MI, USA. His research focuses on MLOps, serverless computing, federated learning, distributed machine learning systems, and cloud-native architectures for large-scale training and inference. He has published in venues including IEEE Access, JSS, ICSOC, AAAI, IEEE IC2E, and IEEE QRS. He received the Ph.D.\ degree in computer science from the Universit\'{e} du Qu\'{e}bec \`, where he investigated serverless architectures for scalable peer-to-peer machine learning training, and the M.Sc.\ degree in software engineering from Polytechnique Montr\'{e}al, where his research applied machine learning techniques to vulnerability detection in source code. Amine also was the recipient of the Tunisian Government Excellence Scholarship and best student paper award at CASCON 2019. 
\end{IEEEbiography}

\end{document}